\documentclass[pra,aps,twocolumn,floatfix]{revtex4}
\usepackage{graphicx}
\usepackage{amsmath}    
\usepackage{amsfonts}   
\usepackage{amssymb}    
\usepackage{amsbsy} 	
\usepackage{hyperref}
\usepackage{epstopdf}
\begin{document}
	
	\title{Observation of Dynamic Stark Resonances in Strong-Field Excitation}
	
	\author{D.~Chetty$^1$} \email{dashavir.chetty@griffithuni.edu.au}
	\author{R.~D.~Glover$^{1,2}$}
	\author{B.~A.~deHarak$^{1,3}$}
	\author{X.~M.~Tong$^{4}$}
	\author{H.~Xu$^1$}
	\author{T.~Pauly$^{5}$}
	\author{N.~Smith$^{5}$}
	\author{K.~R.~Hamilton$^5$}
	\author{K.~Bartschat$^{5}$}
	\author{J.~P.~Ziegel$^3$}
	\author{N.~Douguet$^{6}$}
	\author{A.~N.~Luiten$^2$}
	\author{P.~S.~Light$^2$}
	\author{I.~V.~Litvinyuk$^1$}
	\author{R.~T.~Sang$^1$}	
	\email{r.sang@griffith.edu.au}
	\affiliation{$^1$Centre for Quantum Dynamics, Griffith University, Brisbane, QLD 4111, Australia}
	\affiliation{$^2$Institute for Photonics and Advanced Sensing and School of Physical Sciences, The University of Adelaide, Adelaide, SA 5005, Australia}
	\affiliation{$^3$Physics Department, Illinois Wesleyan University, Bloomington, IL 61702-2900 USA}
	\affiliation{$^4$Center for Computational Sciences, University of Tsukuba, 1-1-1 Tennodai, Tsukuba, Ibaraki 305-8573, Japan}
	\affiliation{$^5$Department of Physics and Astronomy, Drake University, Des Moines, Iowa 50311, United States of America}
	\affiliation{$^6$Department of Physics, Kennesaw State University, Marietta, GA 30060, USA.}

	\date{\today}

\begin{abstract}
We investigate AC Stark-shifted resonances in argon with ultrashort near-infrared pulses. 
    Using 30~fs pulses we observe periodic enhancements of the excitation yield in the intensity 
    regions corresponding to the absorption of 13 and 14 photons. 
    By reducing the pulse duration to 6~fs with only a few optical cycles, we also demonstrate that 
    the enhancements are significantly reduced beyond what is measurable in the experiment. 
    Comparing these to numerical predictions, which are in quantitative agreement with experimental results, 
    we find that even though the quantum-state distribution can be broad, the enhancements are largely due to 
    efficient population of a select few AC Stark-shifted resonant states rather than the closing of an ionization channel. 
    Because these resonances are dependent on the frequency and intensity of the laser field, the broad bandwidth 
    of the 6~fs pulses means that the resonance condition is fulfilled across a large range of intensities. 
    This is further exaggerated by volume-averaging effects, resulting in excitation of the $5g$ state at 
    almost all intensities and reducing the apparent magnitude of the enhancements. 
    For 30~fs pulses, volume averaging also broadens the quantum state distribution but the enhancements are 
    still large enough to survive. In this case, selectivity of excitation to a single state is reduced 
    below 25\% of the relative population. However, an analysis of TDSE simulations indicates that excitation of 
    up to 60\% into a single state is possible if volume averaging can be eliminated and the intensity can be 
    precisely controlled.

\end{abstract}

\maketitle

\section{Introduction}
Strong-field excitation occurs when the interaction of an atom with an intense laser field results in excitation into higher energy states. 
In noble gases, a significant portion of these states decay into long-lived metastable states~\cite{NubPRL08,ZimPRL15}. 
These states have unique properties that enable diverse applications, such as atom lithography~\cite{Baker2004}, 
radiometric dating by way of atom-trap trace analysis~\cite{Lu2014,Sturchio2014}, and precision measurements in beta decay~\cite{Knecht2013,Ohayon2018}. 
In recent years, there has been a demand for higher efficiency and cleaner sources of metastable atoms, encouraging all-optical methods of generation to be pursued. 
Examples include two-photon absorption~\cite{Dakka2018} or methods employing UV lamps~\cite{Kohler2014}. 
Strong-field excitation is also a promising technique. However, 
efficient excitation schemes need to be developed to compete with current metastable-generation methods.

In strong laser fields, excitation rates exhibit a complex dependence on the laser intensity, showing distinct 
enhancements at specific intensities dependent on the target atom~\cite{Li2014,Piraux2017,ZimPRL17}. 
The intense electric field of the laser modifies the energy levels of the atom due to the AC (or dynamic) 
Stark shift~\cite{Delone_1999}, resulting in resonances and thresholds at which excitation 
yields may increase~\cite{Freeman1987,Grasbon2003,Rudenko2004,Kruit1983,Muller1983,Muller1999,Muller1999a,Nandor1999,Muller1998,Li2015}. 
For example, the modification of narrow features in the photo\-electron spectra 
or unexpected changes in the ionization yield at select intensities have been observed and explained through Freeman 
resonances~\cite{Freeman1991,Potvliege2009}, ``channel closing"~\cite{Kruit1983,Muller1983}, and ``population trapping"~\cite{debPRL92,JonPRA93,Morishita2013}.

When the laser frequency, $\omega$, is lower than the frequency of the transition between the ground state and the first excited state, 
the ground-state energy drops by $-\alpha_0I/4$, where $I$ is the laser intensity and $\alpha_0$ is the static polarizability 
of the atom (atomic units are used throughout). The continuum threshold, on the other hand, increases with the intensity-dependent 
ponderomotive energy of the electron, $U_{p}=I/4\omega^2$~\cite{Delone_1999}. Together these shifts
can exceed the energy of a single photon, thus increasing the number of photons required for photo\-ionization from $N$ to $N+1$.  
At this point, the $N$-photon ionization channel is said to close, thereby providing the condition for an $N$-photon channel closing as,
\begin{equation}
    N\hbar\omega = I_{p}+\frac{I}{4}\left( \frac{1}{\omega^2} + \alpha_0 \right),
\end{equation}
where $I_{p}$ is the field-free ionization potential. The AC Stark effect also shifts the energy levels of the excited states. For states with a 
binding energy much less than the ground state, 
this shift closely follows the continuum threshold. Therefore, as the $N$-photon ionization channel closes, high-lying Rydberg states are 
expected to come into resonance. As the intensity increases further, 
lower-lying states will subsequently shift into resonance. 
If these states defy ionization from the remaining cycles of the laser pulse, for example through stabilization~\cite{Volkova2011,Popov2003,Chin2016}, 
their population may accumulate through population trapping. 

In experiments investigating above-threshold ionization, these resonance features in argon photo\-electron spectra were found to strongly 
depend on the laser intensity~\cite{Hertlein1997}. 
Soon after this observation, several theoretical papers were published~\cite{Muller1999a,Nandor1999,Muller1998,Muller2001} detailing that 
the strong intensity dependence is due to low-lying excited states shifting into resonance with $N$-photon absorption. 
Hart \textit{et al.}~\cite{Hart2016} extended this technique to sodium atoms, demonstrating enhanced ionization at a specific intensity that 
corresponds to a Freeman resonance for 3-photon absorption into the Stark-shifted $5p$ state.
These studies, however, did not include the impact on total excitation rates, which is central to the aims of the present investigation. 

A recent experiment demonstrated the resultant impacts by directly observing the excitation yields of argon using 45~fs pulses centered at 400~nm~\cite{ZimPRL17}. 
An increase of more than an order of magnitude was observed at the 6-photon channel closing. The same experiment with 800~nm pulses, however, 
could not resolve any enhancements, even though calculations predict them to persist. 
Extending this, an even more recent experiment~\cite{Hu2019} appeared to resolve these peak structures in strong-field excitation of xenon 
with 50~fs pulses centered at 800~nm. 
In this experiment, a field-ionization technique was employed to detect any excited xenon atoms with principal quantum number $20<n<30$. 
Small features were observed in the ratio of field-ionized neutrals to singly ionized xenon that were attributed to the remainder of the 
peak structure after focal volume averaging. 

In this paper, we present experiments probing strong-field excitation of argon with 30~fs and 6~fs FWHM pulses centered at 800~nm with 
intensities between the multi\-photon and tunneling regimes, remaining below-the-barrier throughout. 
In particular, we focus on the intensities where enhancements are predicted to be most pronounced based on time-dependent Schr\"odinger equation (TDSE) calculations. 
By directly detecting excited states we observe these enhancements experimentally and demonstrate that they are no longer visible for few-cycle pulses. 
The intensities at which these enhancements occur, as well as an analysis of the $nl$ quantum-state distributions predicted by the TDSE, 
show that the enhancements are due to population trapping rather 
than the closing of an ionization channel.

\begin{figure}
	\centering
	\includegraphics[width=\linewidth]{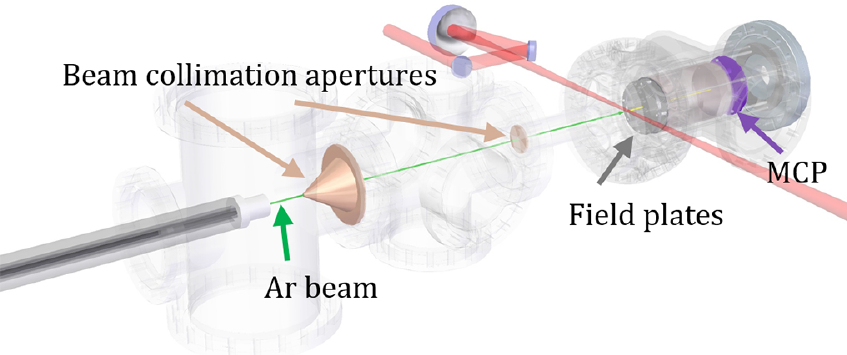}
	\caption{Scheme of the experimental setup. Linearly polarized laser pulses with duration of either 6 or 30~fs (FWHM) centered at 
	800~nm are focused into a collimated effusive argon atomic beam. 
	The atomic beam is collinear with a time-of-flight apparatus backed by a microchannel plate (MCP) that allows the identification of particles. 
	The ions are accelerated and temporally separated from the excited neutrals, which remain at thermal speeds. See text for details.}
	\label{fig:setup}
\end{figure}

\section{Experimental procedure}
We directly detect surviving excited Ar atoms after interacting with ultrashort pulses centered at 800~nm with intensities between 70 and 250~TW/cm$^{2}$. 
The apparatus is depicted in Fig.~\ref{fig:setup}. We use a commercially available (Femto Power) laser system to generate 30~fs pulses. 
Optionally, these pulses can be further compressed using a hollow core fiber to generate 6~fs pulses. 
The intensity is varied by attenuating the pulse energy using a combination of numerous thin membrane pellicle beam-splitters in order 
to preserve the broadband spectrum and chirp of the pulses. 
These are then focused and crossed with a 500~$\mu$m-wide thermal argon atomic beam. A time-of-flight apparatus collinear with the 
atomic beam and a micro-channel plate (MCP) detector are used to discriminate different particles. 
Ions are accelerated by the electric fields and detected within a few tens of microseconds while excited neutral atoms, Ar$^*$, 
remain at thermal speeds and arrive in a 0.15-0.6~ms window. 
These excited states may decay to the long-lived metastable states $(3p^5 4s)^3P_{2,0}$ during the flight and are directly 
detected after Penning ionization on the MCP surface due to their high internal energy ($>$11~eV)~\cite{PenPRL01}. 

\section{Theoretical Methods}
For the numerical simulations, we solve the TDSE in the single-active-electron approximation (SAE) with the model potential given in Ref.~\cite{Tong05c}. 
The radial space is discretized in a generalized pseudo-spectral grid~\cite{Tong97a} and the time-dependent wave function is 
propagated by the second-order split-operator method~\cite{Bandrauk93}. 
We separate the finite box into an inner and outer region to avoid unphysical reflection from the boundary. When the time-dependent 
wave function propagates into the outer region, we project the wave function 
onto momentum space to extract the ionization information and then remove it from the wave function in real space as discussed in~\cite{Tong06b}. 
The final ionization probabilities are obtained by integrating the electron momentum distribution over the entire momentum space.
After the pulse, we project the inner-region wave function on the field-free atomic excited states to get the $nl$ quantum state population up to $n=22,l=21$. 
Summing over all these populations, we obtain the total excitation probability, $P({\rm Ar}^*)$. 

\begin{figure}[t]
	\includegraphics[width=\linewidth]{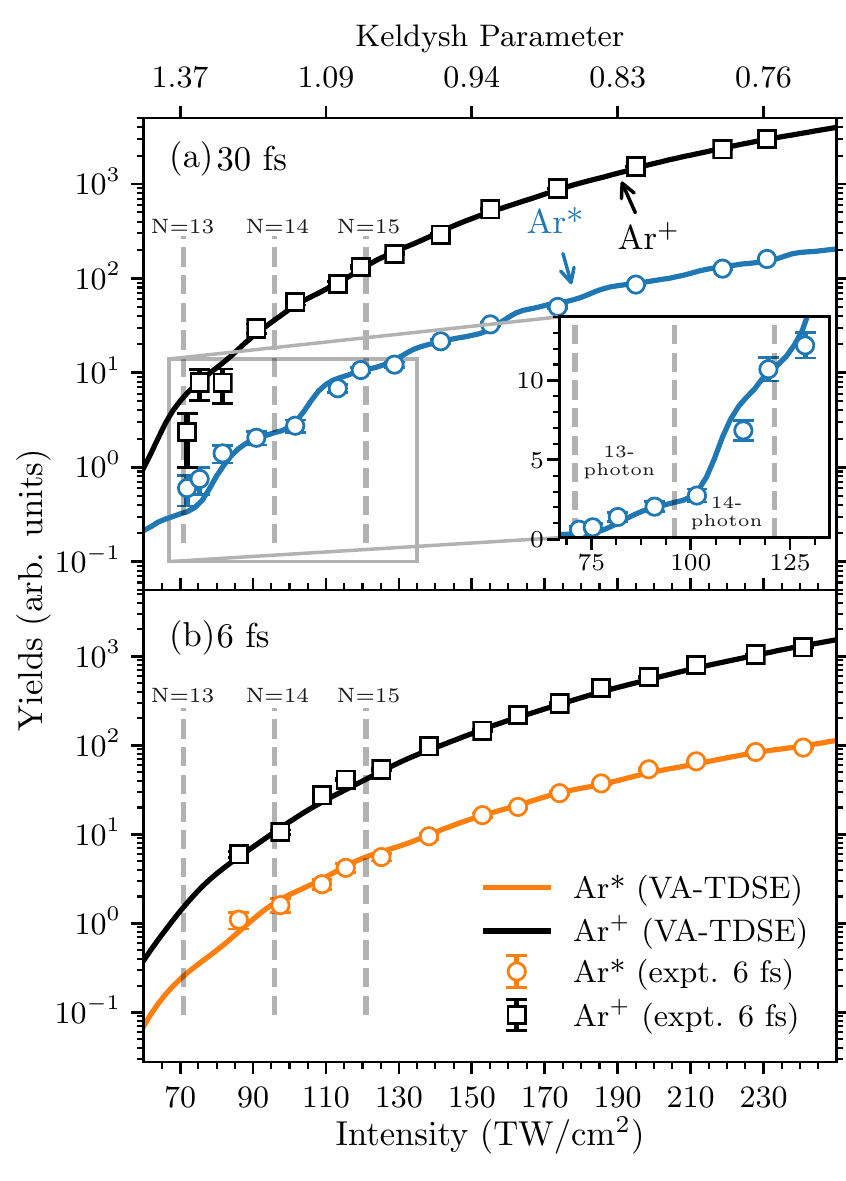}
	\caption{Yields of singly ionized (black) and excited Ar atoms, Ar$^{*}$, as a function of laser intensity for 30~fs~(a) and 6~fs~(b) laser pulses. 
	The solid lines represent the results of the volume-averaged TDSE simulations and include CEP averaging for the 6~fs pulses. 
	The Keldysh parameter is shown above the upper $x$ axis. The zoomed inset shows the region between the 13- and 15-photon 
	channel closings, corresponding to resonances with 13- and 14-photon absorption where a clear modulation is observed for excitation with 30~fs pulses.}
	\label{fig:yieldvsintensity}
\end{figure}

The results from the procedure outlined above was compared to independent calculations~\cite{Gryzlova2018,BROWN2019107062} using the same and 
other similar SAE potentials, such as those suggested in~\cite{GSZ} or generated {\it ab initio} from structure codes like~\cite{SUPERSTRUCTURE}.
The predictions from the various calculations agree to within 5\% at lower intensities and 15\% at higher intensities when the same potential is used. 
As expected, the deviations are somewhat larger for different potentials, but qualitatively the agreement remains satisfactory. 

To compare directly with experiment, we volume average (VA) the theoretical probabilities to account for the intensity distribution around the laser 
focus as in Ref.~\cite{ZimPRL17}. Since the carrier envelope phase~(CEP) of the 6~fs pulse is not stabilized in the experiments, 
the calculations were averaged over four CEP values from 0 to $\pi$ in steps of $\frac{\pi}{4}$.
The experimental intensity for the 6~fs data was calibrated by fitting the ion yield to a phenomenological model~\cite{WallacePRL2016}.
For the 30~fs data, a two-step process is implemented. 
The intensity was initially estimated by fitting the ion yield to the ionization rates predicted by an analytical non-adiabatic model for 
ionization~\cite{Li2016}, resulting in an uncertainty in the intensity of less than 11\%. 
The initial step is necessary to establish an estimated intensity with an uncertainty less than the channel-closing interval.  This allows 
us to align the experimental measured peaks to the correct channel. We then fit the Ar$^*$ yields to the VA-TDSE results~(solid lines 
in Fig.~\ref{fig:yieldvsintensity}) with constrained parameters from step 1 to obtain a more accurate calibrated intensity ($\pm2$\%). 
As a consistency check, this fitting procedure was repeated for ionization rates from the TDSE results.
This produced a calibration factor in agreement with the fit to excitation rates within the uncertainty.
With this method the location of the enhancements provides excellent markers for calibrating the experimental intensity~\cite{ZimPRL17}. 

\section{Results and discussion}
The experimental yields of Ar$^{+}$ (squares) and Ar$^*$ (circles) as a function of the calibrated intensity for 30~fs~(a) and 6~fs~(b) 
pulses are shown in Fig.~\ref{fig:yieldvsintensity}. 
Within the experimental uncertainty, the observed ionization yields exhibit a monotonous increase with increasing intensity. However, with 30~fs pulses, 
some features are clearly visible in the metastable yield, which are washed out for 6~fs pulses. 
We observe good agreement between the experimental data and the VA-TDSE calculations. 
In particular, the features in the Ar$^*$ yields at the 13- and 14-photon absorption channels are well reproduced. 
\begin{figure}
	\includegraphics[width=\linewidth]{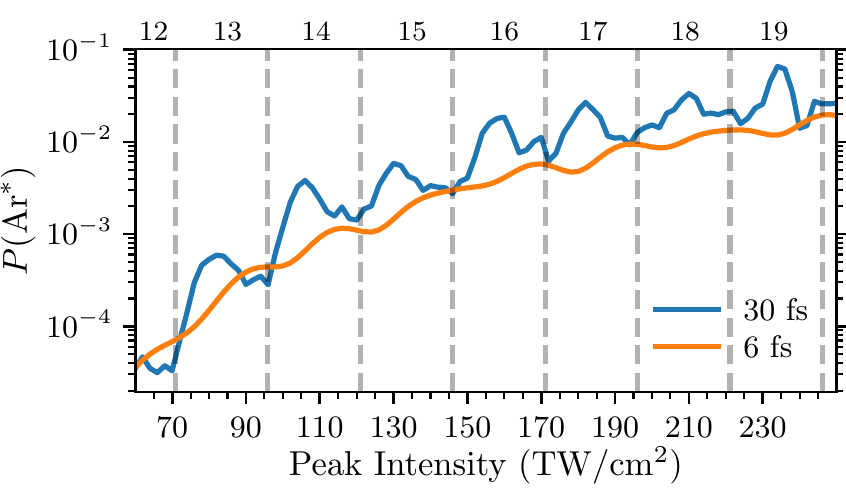}
	\caption{TDSE calculations for the total excited state probability, $P({\rm Ar}^*)$, for 30~fs pulses in blue (darker line) and 6~fs pulses in orange (lighter line) without volume averaging. 
	The numbers above the upper $x$ axis correspond to the number of absorbed photons resulting in excitation within that channel. 
	The dashed lines indicate the intensities at which an ionization channel closes.}
	\label{fig:Prob}
\end{figure}

In order to determine the nature of these features, we further analyze the results from the VA-TDSE calculations (see the Supplementary 
Material for joint $nl$ distributions). 
We note that the features in the 13- and 14-photon absorption channels with 30~fs pulses appear near intensities where the
AC Stark effect shifts the $5g$~(86~TW/cm$^2$) and $6h$~(110~TW/cm$^2$) states into strongest resonance, respectively. However, due to VA effects 
the distribution of quantum states is still relatively broad, with the resonant state accounting for only 17\% and 21\% of the total population.
As a general trend, we see that while the spread of the quantum-state distribution varies widely across intensities, the most populated states remain 
the $6h, 7h$ and $8h$ states from the 14-photon channel onwards.  
Similarly, the spread in quantum state distribution varies for the 6~fs pulses, but the most populated state remains at the $5g$ state for all intensities 
higher than 82~TW/cm$^2$. 
Resonances with some of these states were already predicted (see, for example, Ref.~\cite{Muller2001}), but here we demonstrate that their influence on excitation 
rates are strong enough to be directly measured in our experiment even after VA and experimental instabilities. 
This is further evidence that the AC Stark effect has a significant influence on excitation rates --- not only in regards to channel closings, 
which have been linked to similar features previously, but also due to shifted resonances.   

The VA results include contributions from lower intensities that wash out or obscure patterns, making it difficult to distinguish whether 
channel closings or resonances are the cause of these enhancements. 
The results of the TDSE calculations without VA provide a useful tool for distinguishing these processes and are shown in Fig.~\ref{fig:Prob}. 
The numbers displayed above the upper $x$~axis correspond to the number of absorbed photons required for excitation into that channel. 
Successive channel closings occur at $\sim26$~TW/cm$^{2}$ intervals for 800~nm photons and are marked with vertical dashed lines.  
The general trend is as expected, exhibiting clear enhancements with a periodicity equal to the photon energy separation.
For 30~fs pulses, the enhancements are more pronounced at lower intensities, reaching more than an order of magnitude in the 13- and 14-photon 
absorption channels, consistent to the findings reported in Ref.~\cite{ZimPRL17}. 
These particular enhancements are significant and are observed under our experimental conditions.  For 6~fs pulses, the enhancements are less 
pronounced and not resolved experimentally due to VA effects. 
For both pulse durations, the enhancements occur at higher intensities than the predicted channel closings (at $\sim12$TW/cm$^{2}$ 
and $\sim22$TW/cm$^{2}$ for 30~fs and 6~fs pulses, respectively), indicating that resonances rather than channel closings are the origin of these features. 

\begin{figure}
	\includegraphics[width=\linewidth]{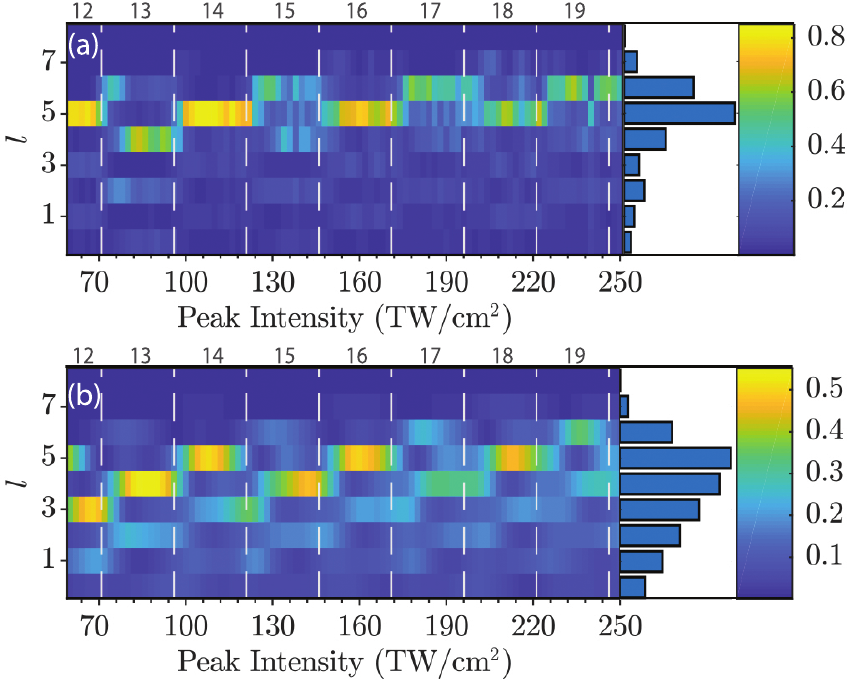}
	\caption{Relative $l$ distributions found by summing over $n\leq 22$ for 30~fs~(a) and 6~fs~(b) pulses without volume averaging. The numbers above the upper $x$ axis 
	correspond to the number of absorbed photons resulting in excitation within that channel. 
	The bar graphs represent the distribution in $l$ summed across all intensities. For both pulse durations, the $l$ distribution 
	clearly alternates between even and odd parity 
	at the closure of successive ionization channels, providing evidence that an additional photon has been absorbed.}
	\label{fig:parity}
\end{figure}

In order to confirm this interpretation, we first validate that channel closings occur at the predicted intensities by analyzing the 
relative populations of the quantum angular momentum, $l$, for each intensity. 
This is done by summing the $nl$ populations over all $n$ and then scaling to the total probability for excitation at that intensity (from Fig.~\ref{fig:Prob}). 
The distribution in $l$ exhibits parity, preferentially exciting even or odd states due to the dipole selection rules~\cite{Krajewska2012}. 
This has been studied previously both semi-classically~\cite{Arbo2008} and quantum mechanically~\cite{Hu2019,Li2014,Piraux2017,Chen2006}. 
In argon, which has a $3p$ ($l=1$) outermost electron in the ground state, the absorption of an even (odd) number of photons will 
preferentially populate odd (even) $l$'s. This is clearly observed in the $l$ distributions 
shown in Fig.~\ref{fig:parity} for both pulse durations, particularly at lower intensities. 
The change in parity at successive channel-closing intensities is consistent with the condition that one more photon is absorbed, 
thus confirming the calculated channel-closing locations. 

Additionally, for 30~fs pulses, we observe that the population distribution is localized with excitation into $l=5$ dominating 
(c.f., the bar graph in Fig.~\ref{fig:parity}). 
For 6~fs pulses, the most populated states remain at $l=5$, but now the distribution is broadened by excitation into lower $l$ states. 

\begin{figure}
	\includegraphics[width=\linewidth]{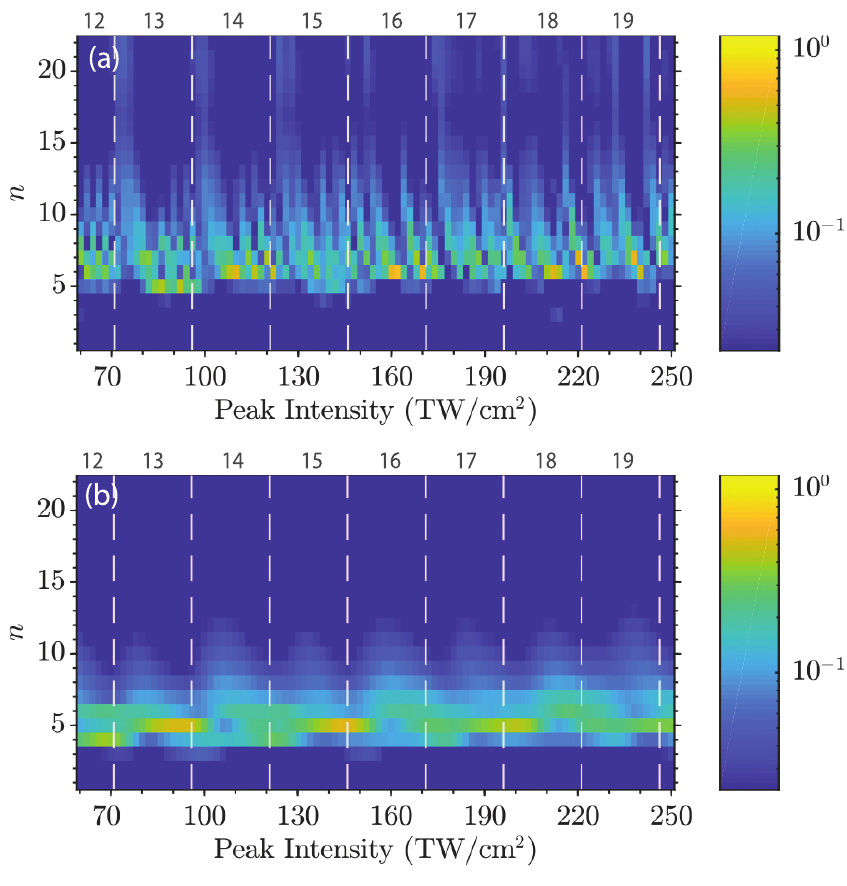}
	\caption{Relative $n$ populations for 30~fs~(a) and 6~fs~(b) pulses without volume averaging. 
	The numbers above the upper $x$ axis correspond to the number of absorbed photons 
	resulting in excitation within that channel. The dashed lines indicate the intensities at which an ionization channel closes. 
	High $n$ states are excited at the channel closing intensities, 
	shifting to individual resonances with the $6h$ (for 30~fs pulses) and $5g$ state (for both) as the intensity is increased further.} 
	\label{fig:n-dist}
\end{figure}

We now look to at the relative $n$ populations to analyze the patterns around channel closings. 
These are obtained in a similar procedure as the relative $l$ populations, except by summing 
over $l$ rather than $n$; see Fig.~\ref{fig:n-dist}. 
In addition, we correlate these observations with those in Fig.~\ref{fig:parity} for a complete 
description of the excited-state distribution. See also the supplementary material for joint $nl$ distributions. 
For 30~fs pulses~(c.f.\ Fig.~\ref{fig:n-dist}(a)), a broad range of high-lying excited states~($n\geq 12$) 
is populated shortly after the channel-closing intensity as the AC Stark effect shifts the Rydberg quasicontinuum into resonance. 
For 6~fs pulses (c.f.\ Fig.~\ref{fig:n-dist}(b)) the pattern is much the same but not as obvious. 
This is because the pulse duration is now short compared to the Keppler orbit periods of high-lying Rydberg states, 
which are therefore not populated as efficiently~\cite{Grasbon2003}.
As the intensity increases further, the distribution narrows until a strong resonance with either the $5g$ 
(for both pulse durations) or $6h$ (for 30~fs) state is reached. 

The behavior of this resonance is markedly different for the two pulse durations. 
Firstly, the intensities at which the strongest resonances are reached in successive channels are different. 
For example, with 30~fs pulses, the strongest resonance in the 13- and 14-photon absorption channels is 
reached with the $5g$ and $6h$ states at $86$~TW/cm$^{2}$ and 110~TW/cm$^{2}$, respectively. On the other hand, 
with 6~fs pulses it is reached at $90$~TW/cm$^{2}$ and 122~TW/cm$^{2}$. 
Secondly, the resonances are less dominant and occur over a wider range of intensities for 6~fs pulses 
compared to 30~fs pulses due to the larger bandwidth enabling resonances over a wider range of photon energies.
For example, with 30~fs pulses at 162~TW/cm$^{2}$, the $6h$ state accounts for almost 60\% of total excitation 
but then drops close to zero only 4~TW/cm$^{2}$ higher. 
In comparison, for 6~fs pulses, resonance with the $5g$ state occurs in a 12~TW/cm$^{2}$ intensity range accounting 
for over 30\% of relative population, peaking at 146~TW/cm$^{2}$ with 35\% relative population. 
This reduced dominance, as well as the larger intensity range where resonance is reached, accounts for the reduced 
magnitude of the enhancements. 

Interestingly, we note that even though the intensities of these strong individual resonances are very close 
to those corresponding to the enhancements observed in the measurements~(Fig.~\ref{fig:yieldvsintensity}) 
and theoretical yields~(Fig.~\ref{fig:Prob}), they are not the sole contributors. 
A detailed analysis of the joint $nl$ distributions from 30~fs pulses indicates that the main contributions to the 
peaks of the 13- and 14-photon enhancements originate from AC Stark-shifted resonances with a trio of states with 
successive $n$ and same $l$ ($5g$, $6g$, $7g$ and $6h$, $7h$, $8h$, respectively). 
In the case of 6~fs pulses, excitation into the $5g$ state mainly contributes to the enhancements in odd photon 
channels, while a broad distribution contributes to the observed enhancements in even photon channels, at least 
in the multi\-photon regime where the locations of the enhancements are obvious. 

\section{Summary}

We experimentally observed enhancements in excitation rates of Ar for 30~fs pulses centered at 800~nm, 
which were not present for few-cycle pulses of 6~fs duration. 
TDSE calculations support the existence of these enhancements even after focal-volume averaging. 
Due to the sensitivity of these enhancements to intensity changes, they serve as convenient markers 
for accurate calibration of the experimental intensity. 
Analysis of the TDSE predictions shows that the enhancements are due to resonant population trapping 
in select few states rather than the closing of an ionization channel. 
Volume averaging effects suppress the relative populations of these states at resonant intensities. 
However, TDSE calculations predict that the resonances are particularly strong for select intensities 
when using 30~fs pulses but spread over a larger intensity range for 6~fs pulses due to the large bandwidth of the pulse.  
In future, enhanced excitation of the $5g$ and $6h$ states might be exploited as a means to increase 
metastable yields by directly stimulating them into the metastable state. 

\section*{ACKNOWLEDGEMENTS}
This project is supported under the Australian Research Council's Linkage Infra\-structure, 
Equipment and Facilities scheme (project LE160100027).  
D.~Chetty is supported by an Australian Government RTP Scholarship. 
X-.M.~T.~was supported by a Grant-in-Aid for Scientific Research (Grant No.\ JP16K05495) from 
the Japan Society for the Promotion of Science.
Further support was provided by the United States National Science Foundation under grants 
No.\ PHY-1402899 and PHY-1708108 (BdH,JPZ) as well as No.\ PHY-1803844 (TP,NS,KRH,KB).

\bibliography{FrustratedBib}
\end{document}